\documentclass[aps,pre,onecolumn,groupedaddress,longbibliography,nofootinbib,showkeys,showpacs,amssymb]{revtex4-1}

\usepackage{latexsym}
\usepackage{amsfonts}
\usepackage{graphicx}
\usepackage{mathptmx}      
\usepackage{bm}
\usepackage{enumerate}
\usepackage{subfigure}

\usepackage{color}

\DeclareRobustCommand\openone{\leavevmode\hbox{\small1\normalsize\kern-.33em1}}

\newcommand\unitS{{a}}

\begin{document}


\title{Quantum theory as a description of robust experiments:\\
derivation of the Pauli equation\footnote{Accepted for publication in Ann. Phys.}}



\author{Hans De Raedt}
\affiliation{Department of Applied Physics, Zernike Institute for Advanced Materials,\\
University of Groningen, Nijenborgh 4, NL-9747AG Groningen, The Netherlands}
\author{Mikhail I. Katsnelson}
\affiliation{Radboud University Nijmegen, Institute for Molecules and Materials,
Heyendaalseweg 135, NL-6525AJ Nijmegen, The Netherlands}
\author{Hylke C. Donker}
\affiliation{Radboud University Nijmegen, Institute for Molecules and Materials,
Heyendaalseweg 135, NL-6525AJ Nijmegen, The Netherlands}
\author{Kristel Michielsen}
\thanks{Corresponding author}
\email{k.michielsen@fz-juelich.de}
\affiliation{Institute for Advanced Simulation, J\"ulich Supercomputing Centre,\\
Forschungszentrum J\"ulich, D-52425 J\"ulich, Germany}
\affiliation{RWTH Aachen University, D-52056 Aachen, Germany}

\date{\today}

\begin{abstract}
It is shown that the Pauli equation and the concept of spin naturally emerge
from logical inference applied to experiments on a charged particle
under the conditions that
(i) space is homogeneous
(ii) the observed events are logically independent, and
(iii) the observed frequency distributions are robust with
respect to small changes in the conditions under which the experiment is carried out.
The derivation does not take recourse to concepts of quantum theory and is based on the same principles
which have already been shown to lead to e.g. the Schr\"odinger equation
and the probability distributions of pairs of particles in the singlet or triplet state.
Application to Stern-Gerlach experiments with chargeless, magnetic particles,
provides additional support for the thesis that quantum theory follows from
logical inference applied to a well-defined class of experiments.
\end{abstract}


\maketitle


{\bf Highlights}
\begin{itemize}
\item
The Pauli equation is obtained through logical inference applied to robust experiments on a charged particle.
\item
The concept of spin appears as an inference resulting from the treatment of two-valued data.
\item
The same reasoning yields the quantum theoretical description of neutral magnetic particles.
\item
Logical inference provides a framework to establish a bridge between objective knowledge gathered
through experiments and their description in terms of concepts.
\end{itemize}

\section{Introduction}\label{LI}

In laboratory experiments, one {\bf never} has complete knowledge about the
mechanisms that affect the outcome of the measurements: there is always uncertainty.
In addition, the outcomes of real experiments are always subject to uncertainties
with respect to the conditions
under which the experiments are carried out.

If there are uncertainties about the individual events
and uncertainties about the conditions under which the experiment is carried out,
it is often difficult or even impossible to establish relations between individual events.
However, in the case that the frequencies of these events
are robust (to be discussed in more detail later) it may still be possible
to establish relations, not between the individual events,
but between the frequency distributions of the observed events.

The algebra of logical inference provides a mathematical framework that
facilitates rational reasoning when there is uncertainty~\cite{COX46,COX61,TRIB69,SMIT89,JAYN03}.
A detailed discussion of the foundations of logical inference, its relation to Boolean logic and the derivation of
its rules can be found in the papers~\cite{COX46,SMIT89} and books~\cite{COX61,TRIB69,JAYN03}.
Logical inference is the foundation for powerful tools such as the maximum entropy method
and Bayesian analysis~\cite{TRIB69,JAYN03}.
To the best of our knowledge, the first derivation of a non-trivial theoretical description by
this general methodology of scientific reasoning appears
in Jaynes' papers on the relation between information and (quantum) statistical mechanics~\cite{JAYN57a,JAYN57b}.

A recent paper~\cite{RAED14b} shows how some of the most basic equations of quantum theory,
e.g. the Schr\"odinger equation and the probability distributions of
pairs of particles in the singlet or triplet state
emerge from the application of logical inference to (the abstraction of) robust experiments,
without taking recourse to concepts of quantum theory.
This logical-inference approach yields results that are unambiguous and independent of the individual subjective judgement.
In addition, this approach provides a rational explanation
for the extraordinary descriptive power of quantum theory~\cite{RAED14b}.
As the introduction of the concept of intrinsic angular momentum, called spin,
is a landmark in the development of quantum theory, it is natural
to ask the question under which circumstances this concept appears in a logical-inference treatment.

A classical review of how the concept of spin has been introduced
in quantum theory is given by van der Waerden~\cite{WAER60}.
The original motivation to introduce this new concept was the discovery of the anomalous Zeeman effect
and its transition to the normal Zeeman effect with increasing magnetic field (the so-called Paschen-Back effect).
Pauli introduced spin in a very formal way by attributing to the electron
an additional intrinsic magnetic quantum number taking the values $\pm1/2$~\cite{PAUL24}.
Although the picture of the spin in terms of a ``rotating electron model''
was quickly and widely accepted, Pauli was strongly against this picture because of its purely classical-mechanics character.
A few years later he suggested the Pauli equation~\cite{PAUL27}
in which this intrinsic degree of freedom was introduced by replacing
the single-component wavefunction that appears in Schr\"odinger's equation
by a two-component wavefunction and ``Pauli matrices'';
the most rigorous way to establish a relation with the idea of the rotating electron
is just a formal observation that these Pauli matrices satisfy the same commutation
relation as the generators of the rotation group in three-dimensional space
and that the two-component wavefunctions (spinors) provide a double-valued representation of this group~\cite{WAER60}.

Bohm and his followers,
in the spirit of their general approach to provide a causal interpretation of quantum mechanics,
tried to construct a purely classical description of spin
by analogy with the hydrodynamics of a rotating liquid~\cite{BOHM55a,TAKA55}.
Despite the beauty of the mathematical description, the interpretation
of the spin as entity, a field, which is distributed over the whole space
is rather exotic and can hardly be considered as a derivation and justification of the Pauli equation.

Bohr and Pauli suggested that spin and the related magnetic moment cannot be measured
in experiments which can be interpreted in terms of classical trajectories
(such as Stern-Gerlach experiments with a free-electron beam),
see Ref.~\onlinecite{BATE97} and references therein.
In an inhomogeneous magnetic field, spin effects
cannot be separated from the effects of the Lorentz force due to the orbital motion of the charged particle.
However, these difficulties are technical rather than conceptual
as they do not consider the possibility that there are neutral particles (not subject to the Lorentz force)
with magnetic moments, such as neutrons, for which Stern-Gerlach experiment is not only possible in principle
but has really been performed~\cite{SHER54}.
It is clear now that the naive way
to demonstrate the ``essentially non-classical'' character of the spin degree of freedom premature.

In this paper, we show how the Pauli equation and the concept of spin naturally emerge
from the logical-inference analysis of experiments on a charged particle.
We carefully analyze the additional assumptions (some of them
having obvious analogs in Pauli's analysis of the anomalous Zeeman effect)
which are required to pass, in a model-free way, to the Pauli equation.

Conceptually, we return to the roots by first introducing
``spin'' as some intrinsic degree of freedom characterized by a two-valued number.
We will call this two-valued property ``color'' (e.g. red or blue)
to make clear that we leave no room for
(mis)interpretations in terms of models of rotating particle and the like.
This is in sharp contrast to the interpretation of Refs.~\onlinecite{BOHM55a,TAKA55}.
Note that such a generalization of the concept of spin is very important in modern physics.
For instance, the idea of isospin of elementary particles~\cite{GRIF87}
which was originally introduced~\cite{HEIS32}
as a way to describe constituents of atomic nuclei
in terms of the same particles (nucleons) with two subspecies (neutrons and protons).
Another example is the pseudospin of the charge carriers in graphene~\cite{KATS12}
used to indicate that the carriers belongs to sublattice A or B of the honeycomb crystal lattice.
In both of these examples, there is nothing that is rotating!

We further illustrate the power of the approach by an application
to Stern-Gerlach experiments with chargeless, magnetic particles,
providing additional support to the idea that quantum theory directly follows from
logical inference applied to a well-defined class of experiments~\cite{RAED14b}.

To head off possible misunderstandings, it is important to mention
that the underlying premise of our approach is that
current scientific knowledge derives, through cognitive processes in the human brain, from the discrete events which
are observed in laboratory experiments and from relations between those events that we, humans, discover.
As a direct consequence of this underlying premise, the validity of the results obtained
in our approach does not depend on the assumption that the observed events are signatures
of an underlying objective reality which is mathematical in nature
(for an overview of older and new work in this direction, see Ref.\onlinecite{KHRE14}).
We take the point of view that the aim of physics is to provide a consistent description of
relations between certain events that we perceive (usually with the help of some equipment) with our senses.
Some of these relations express cause followed by an effect and others do not.
A derivation of a quantum theoretical description from logical-inference principles
does not prohibit the construction of cause-and-effect
mechanisms that, when analyzed in the same manner as in real experiments, create the {\it impression}
that the system behaves according to quantum theory~\cite{THOO97,RAED05b,THOO07}.
Work in this direction has shown that it is indeed possible to build simulation models
which reproduce, on an event-by-event basis, the results of interference/entanglement/uncertainty
experiments with photons/neutrons~\cite{MICH11a,RAED12a,RAED12b,RAED14a,MICH14a}.

The paper is organized as follows.
In Section~\ref{LIP} we specify the measurement scenario,
introduce the inference-probability that characterizes the observed detection events
(all the elements of logical inference that are required to
for the purpose of the present paper are summarized in Appendix~\ref{sec2}).
Then, we discuss and formalize the notion of a robust experiment.
Although these three steps are similar to the ones taken
in the logical-inference derivation of the Schr\"odinger equation~\cite{RAED14b},
to make the presentation self-contained, we give a detailed account.
The next three subsections address the problem of including additional knowledge
about the motion of the particle in some limiting cases.
In subsection~\ref{DPE} we collect the results of the previous subsections
and derive the Pauli equation.
Section~\ref{SGE} shows that the same procedure leads to the quantum theoretical equation
that describes the motion of an uncharged particle in a magnetic field.
A discussion of the relation of the logical-inference derivation
of the Pauli equation and earlier work on the hydrodynamic formulation
of quantum theory is given in Section~\ref{EAR}.
A summary and discussion of more general aspects of the work
presented in this paper can be found in Section~\ref{CON}.

\begin{widetext}

\section{Logical inference: derivation of the Pauli equation}\label{LIP}

\subsection{Measurement scenario}

We consider $N$ repetitions of an experiment
on a particle located in $3$-dimensional space $\bm\Omega$.
The experiment consists of sending a signal to the particle
at discrete times labeled by the integer $\tau=1,\ldots,M$.
It is assumed that for each repetition, labeled by $n=1,\ldots,N$,
the particle is at the unknown position $\mathbf{X}_\tau\in \bm\Omega$.
As the particle receives the signal, it responds by emitting
another signal which is recorded by an array of detectors.
For each signal emitted by a particle
the data recorded by the detector system
is used to determine the position $\mathbf{j}_{n,\tau}\in {\cal V}$
where ${\cal V}$ denotes the set of voxels with linear extent $[-\Delta,\Delta]/2$
that cover the $3$-dimensional space $\bm\Omega$.
The signal also contains additional information which is two-valued
and encodes, so to speak, the ``color'' of the particle
at the time when it responded to the signal emitted by the source.
This color is represented by variables $k_{n,\tau}=\pm1$.
The frequency distribution of the
$(\mathbf{j},k)_{n,\tau}$'s changes with the applied electric and magnetic field
from which we may infer that there is some form of interaction between the electromagnetic field and the particle.

The result of $N$ repetitions of the experiment yields the data set
\begin{equation}
\Upsilon=\{(\mathbf{j},k)_{n,\tau}| \mathbf{j}_{n,\tau}\in {\cal V};\; k=\pm1;\; n=1,\ldots,N;\; \tau=1,\ldots,M \}
,
\label{LIa0}
\end{equation}
or, denoting the total counts of voxels $\bm{j}$ and color $k$
at time $\tau$ by $0\le c_{\bm{j},k,\tau}\le N$,
the data set can be represented as
\begin{eqnarray}
{\cal D}&=&\Bigl\{ c_{\mathbf{j},k,\tau}\Bigl| \tau=1,\ldots,M\;; \sum_{k=\pm1} \sum_{\bm{j}\in [-L^d,L^d]} c_{\mathbf{j},k,\tau}=N \Bigr\}
.
\label{LIa1}
\end{eqnarray}

\subsection{Inference-probability of the data produced by the experiment}\label{TDSEb}

The first step is to introduce a real number $0\le P(\mathbf{j},k|\mathbf{X}_\tau,\tau,Z)\le1$ which represents the plausibility that
we observe a detector click $(\mathbf{j},k)$, conditional on $(\mathbf{X}_\tau,\tau,Z)$.
For reasons explained in Appendix~\ref{sec1}, $P(\mathbf{j},k|\mathbf{X}_\tau,\tau,Z)$
is called inference-probability (or i-prob for short) and
encodes the relation between the unknown location $\mathbf{X}_\tau$
and the location $\mathbf{j}$ and color $k$ registered by the detector system
at discrete time $\tau$.
Except for the unknown location $\mathbf{X}_\tau$, all
other experimental conditions are represented by $Z$ and are assumed
to be fixed and identical for all experiments.
Note that unlike in the case of parameter estimation,
in the case at hand both $P(\mathbf{j},k|\mathbf{X}_\tau,\tau,Z)$ and the parameters $\mathbf{X}_\tau$ are unknown.

We make the following, seemingly reasonable assumptions:
\begin{enumerate}[1.]
\item
Each repetition of the experiment represents an event of which
the outcome is logically independent of any other such event.
By application of the product rule (see Appendix~\ref{sec1}), a direct consequence of this assumption is that
\begin{eqnarray}
P(\Upsilon|\mathbf{X}_1,\ldots,\mathbf{X}_M,N,Z)
&=& \prod_{\tau=1}^M\prod_{n=1}^N P(\mathbf{j}_{n,\tau},k_{n,\tau}|\mathbf{X}_\tau,\tau,Z)
,
\label{TDSE2a}
\end{eqnarray}
and hence
\begin{eqnarray}
P({\cal D}|\mathbf{X}_1,\ldots,\mathbf{X}_M,N,Z)
&=& N!\prod_{\tau=1}^M\prod_{\mathbf{j}\in{\cal V}}\prod_{k=\pm1} \frac{P(\mathbf{j},k|\mathbf{X}_\tau,\tau,Z)^{c_{\mathbf{j},k,\tau}}}{c_{\mathbf{j},k,\tau}!}
.
\label{TDSE2}
\end{eqnarray}
\item
It is assumed that it does not matter where the experiment is carried out.
This implies that the i-prob should have the property
\begin{equation}
P(\mathbf{j},k|\mathbf{X}_\tau,\tau,Z)=
P(\mathbf{j}+\bm\zeta,k|\mathbf{X}_\tau+\bm\zeta,\tau,Z)
,
\label{TDSE4}
\end{equation}
where $\bm\zeta$ is an arbitrary $3$-dimensional vector.
The relation Eq.~(\ref{TDSE4}) expresses the assumption that space is homogeneous.
\end{enumerate}

\subsection{Condition for reproducibility and robustness}\label{TDSEc}

If the frequencies with which the detectors fire vary erratically with $\{\mathbf{X}_\tau\}$,
the experiment would most likely be called ``irreproducible''.
Excluding such experiments, it is desirable that frequency distributions of the data
exhibit some kind of robustness, smoothness with respect to small changes of the unknown values
of $\{\mathbf{X}_\tau\}$.
Unless the experimental setup is sufficiently ``robust'' in the sense
just explained, repeating the run with slightly different values of $\{\mathbf{X}_\tau\}$
would often produce results that are very different from those of other runs
and it is common practice to discard such experimental data.
Therefore, a ``good'' experiment must be a robust experiment.

The robustness with respect to small variations
of the conditions under which the experiment is carried out
should be reflected in the expression of the i-prob to observe data sets which yield reproducible
averages and correlations (with the usual statistical fluctuations).
The next step therefore is to determine the expression for
$P(\mathbf{j},k|\mathbf{X}_{\tau},\tau,Z)$ which is most insensitive to small changes in $\mathbf{X}_\tau$.
It is expedient to formulate this problem as an hypothesis test.
Let $H_0$ and $H_1$ be the hypothesis that the same data ${\cal D}$ is observed for
the unknown locations $\{\mathbf{X}_\tau\}$ and
$\{\mathbf{X}_\tau+\bm\epsilon_\tau\}$, respectively.
The evidence $\mathrm{Ev}$ of hypothesis $H_1$, relative to hypothesis $H_0$,
is defined by~\cite{TRIB69,JAYN03}
\begin{eqnarray}
\mathrm{Ev}&=&\ln
\frac{
P({\cal D}|\mathbf{X}_\tau+\bm\epsilon_\tau,\tau,N,Z)
}{
P({\cal D}|\mathbf{X}_\tau,\tau,N,Z)
}
\nonumber \\
&=&
\sum_{\mathbf{j},k,\tau}
c_{\mathbf{j},k,\tau}
\ln\frac{P(\mathbf{j},k|\mathbf{X}_{\tau}+\bm\epsilon_{\tau},\tau,Z)}{P(\mathbf{j},k|\mathbf{X}_{\tau},\tau,Z)}
,
\label{robu1}
\end{eqnarray}
where the logarithm serves to facilitate algebraic manipulations.
If $H_1$ is more (less) plausible than $H_0$ then $\mathrm{Ev}>0$ ($\mathrm{Ev}<0$).
In statistics, the r.h.s. of Eq.~(\ref{robu1}) is known as the log-likelihood function
and used for parameter estimation. In contrast, in the present context,
the function Eq.~(\ref{robu1}) is {\it not} used to estimate
$\mathbf{X}_\tau$ but is a vehicle to express
the robustness with respect to the coordinates $\mathbf{X}_\tau$.

Writing Eq.~(\ref{robu1}) as a Taylor series in $\bm\epsilon$ we have
\begin{eqnarray}
\mathrm{Ev}&=&
\sum_{\mathbf{j},k,\tau}
c_{\mathbf{j},k,\tau}\ln
\left[1+
\frac{\bm\epsilon_\tau\cdot\bm\nabla_\tau P(\mathbf{j},k|\mathbf{X}_{\tau},\tau,Z)}{P(\mathbf{j},k|\mathbf{X}_{\tau},\tau,Z)}
+
\frac{1}{2}\frac{(\bm\epsilon_\tau\cdot\bm\nabla_\tau)^2 P(\mathbf{j},k|\mathbf{X}_{\tau},\tau,Z)}{P(\mathbf{j},k|\mathbf{X}_{\tau},\tau,Z)}
+{\cal O}(\epsilon_\tau^3)
\right]
\nonumber \\
&=&
\sum_{\mathbf{j},k,\tau}
c_{\mathbf{j},k,\tau}\left[
\frac{\bm\epsilon_\tau\cdot\bm\nabla_\tau P(\mathbf{j},k|\mathbf{X}_{\tau},\tau,Z)}{P(\mathbf{j},k|\mathbf{X}_{\tau},\tau,Z)}
-
\frac{1}{2}\left[\frac{\bm\epsilon_\tau\cdot\bm\nabla_\tau
P(\mathbf{j},k|\mathbf{X}_{\tau},\tau,Z)}{P(\mathbf{j},k|\mathbf{X}_{\tau},\tau,Z)}\right]^2
+
\frac{1}{2}\frac{(\bm\epsilon_\tau\cdot\bm\nabla_\tau)^2 P(\mathbf{j},k|\mathbf{X}_{\tau},\tau,Z)}{P(\mathbf{j},k|\mathbf{X}_{\tau},\tau,Z)}
\right]
+{\cal O}(\bm\epsilon_\tau^3)
,
\label{robu3}
\end{eqnarray}
where $\bm\nabla_\tau$ differentiates with respect to $\mathbf{X}_{\tau}$.
Here and in the following we assume that $\bm\epsilon_\tau$ is sufficiently
small such that the third and higher order terms in the $\bm\epsilon$'s can be ignored.
According to our criterion of robustness,
the evidence Eq.~(\ref{robu3}) should change as little as possible as $\mathbf{X}_\tau$ varies.
This can be accomplished by minimizing, in absolute value, all the coefficients
of the polynomial in $\bm\epsilon_\tau$, {\it for all allowed} $\bm\epsilon_\tau$
and $\mathbf{X}_\tau$.
The clause ``for all allowed $\bm\epsilon_\tau$ and  $\mathbf{X}_\tau$'' implies that we are
dealing here with an instance of a global optimization problem~\cite{WIKIROBUST}.

The first and third sum in Eq.~(\ref{robu3})
vanish identically if we choose $c_{\mathbf{j},k,\tau}/N=P(\mathbf{j},k|\mathbf{X}_{\tau},\tau,Z)$.
Indeed, we have
\begin{eqnarray}
\sum_{\mathbf{j},k,\tau}
c_{\mathbf{j},k,\tau}
\frac{(\bm\epsilon_\tau\cdot\bm\nabla_\tau)^\alpha P(\mathbf{j},k|\mathbf{X}_{\tau},\tau,Z)}{P(\mathbf{j},k|\mathbf{X}_{\tau},\tau,Z)}
&=&
N\sum_{\mathbf{j},k,\tau}
(\bm\epsilon_\tau\cdot\bm\nabla_\tau)^\alpha P(\mathbf{j},k|\mathbf{X}_{\tau},\tau,Z)
\nonumber \\
&=&
N\sum_{\tau}
(\bm\epsilon_\tau\cdot\bm\nabla_\tau)^\alpha
\sum_{\mathbf{j},k}
 P(\mathbf{j},k|\mathbf{X}_{\tau},\tau,Z)
\nonumber \\
&=&
N\sum_{\tau}
(\bm\epsilon_\tau\cdot\bm\nabla_\tau)^\alpha
1
=0
,
\label{robu3a}
\end{eqnarray}
for $\alpha=1,2,\ldots$.
Although this choice is motivated by the desire to eliminate contributions of order $\bm\epsilon_\tau$,
it follows that our criterion of robustness automatically
suggests the intuitively obvious procedure to assign to
$P(\mathbf{j},k|\mathbf{X}_{\tau},\tau,Z)$
the value of the observed frequencies of occurrences $c_{\mathbf{j},k,\tau}/N$~\cite{TRIB69,JAYN03}.

Dropping irrelevant numerical factors and terms of ${\cal O}(\epsilon_\tau^3)$,
the remaining contribution to the evidence
\begin{eqnarray}
\mathrm{Ev}&=&
\sum_{\mathbf{j},k,\tau}
\frac{1}{P(\mathbf{j},k|\mathbf{X}_{\tau},\tau,Z)}
\left[\bm\epsilon\cdot\bm\nabla_\tau P(\mathbf{j},k|\mathbf{X}_{\tau},\tau,Z)\right]^2
,
\label{robu4}
\end{eqnarray}
vanishes identically (for all $\bm\epsilon_\tau$) if and only if
$\bm\nabla_\tau P(\mathbf{j},k|\mathbf{X}_{\tau},\tau,Z)=0$
in which case it is clear that we
can only describe experiments for which the data does not exhibit any dependence on $\mathbf{X}_\tau$.
\begin{center}
\framebox{
\parbox[t]{0.9\hsize}{%
Experiments which produce frequency distributions that do not depend on
the conditions do not increase our knowledge about the relation
between the conditions and the observed data.
Therefore, we explicitly exclude such non-informative experiments.
}}
\end{center}
Thus, from now on, we explicitly exclude the class of experiments for which
$\bm\nabla_\tau P(\mathbf{j},k|\mathbf{X}_{\tau},\tau,Z)=0$.

The clause ``for all allowed $\bm\epsilon_\tau$'' can be eliminated using the Cauchy-Schwarz inequality.
We have
\begin{eqnarray}
\mathrm{Ev}
&=&
\sum_{\mathbf{j},k,\tau}
\left[
\frac{\bm\epsilon_\tau\cdot\bm\nabla_\tau P(\mathbf{j},k|\mathbf{X}_{\tau},\tau,Z)}{P^{1/2}(\mathbf{j},k|\mathbf{X}_{\tau},\tau,Z)}
\right]^2
\le
\widehat \epsilon^2
\sum_{\mathbf{j},k,\tau}
\frac{1}{P(\mathbf{j},k|\mathbf{X}_{\tau},\tau,Z)}
\bigr[\bm\nabla_\tau P(\mathbf{j},k|\mathbf{X}_{\tau},\tau,Z)\bigr]^2
,
\label{robu4a}
\end{eqnarray}
where $\widehat \epsilon^2=\max_\tau \bm\epsilon_\tau^2$.
As the $\bm\epsilon_\tau$'s are arbitrary (but small), it follows
from Eq.~(\ref{robu4a}) that we find the robust solution(s)
by searching for the global minimum of
\begin{eqnarray}
I_F
&=&
\sum_{\mathbf{j},k,\tau}
\frac{1}{P(\mathbf{j},k|\mathbf{X}_{\tau},\tau,Z)}
\bigr[\bm\nabla_\tau P(\mathbf{j},k|\mathbf{X}_{\tau},\tau,Z)\bigr]^2
,
\label{robu4b}
\end{eqnarray}
which is the Fisher information of the measurement scenario described above.

\subsection{Continuum limit}\label{two.d}

Propositions such as ``detector $(\mathbf{j},k)$ has clicked at time $\tau$''
are ultimately related to sensory
experience and are therefore discrete in nature.
On the other hand, the basic equations of quantum theory such as the Schr\"odinger, Pauli and Dirac equations
are formulated in continuum space.
Taking the continuum limit of the discrete formulation connects the two modes of description.
Here and in the following, we use the symbols for (partial) derivatives
for both the case that the continuum approximation is meaningful and
the case that it is not. In the latter, operator symbols such as $\partial /\partial t$ should
be read as the corresponding finite-difference operators.

Assuming that the continuum limit is well-defined,
we have ${\cal V}\rightarrow \bm\Omega$ and
the Fisher information reads
\begin{eqnarray}
I_F
&=& \int d\mathbf{x} \, dt
\sum_{i=1}^3
\sum_{k=\pm1}
\frac{1}{P(\mathbf{x},k|\mathbf{X},t,Z)}\left[\frac{\partial P(\mathbf{x},k|\mathbf{X},t,Z)}{\partial X_i}\right]^2
\nonumber \\
&=& \int d\mathbf{x} \, dt
\sum_{i=1}^3
\sum_{k=\pm1}
\frac{1}{P(\mathbf{x},k|\mathbf{X},t,Z)}\left[\frac{\partial P(\mathbf{x},k|\mathbf{X},t,Z)}{\partial x_i}\right]^2
\nonumber \\
&=& \int d\mathbf{x} \, dt
\sum_{k=\pm1}
\frac{1}{P(\mathbf{x},k|\mathbf{X},t,Z)}\left[\bm\nabla P(\mathbf{x},k|\mathbf{X},t,Z)\right]^2
,
\label{LI0}
\end{eqnarray}
where $\bm\nabla$ denotes derivatives with respect to $\mathbf{x}$
and we have simplified the notation somewhat by writing $\mathbf{X}=\mathbf{X}_t$.
We have changed derivatives with respect to $\mathbf{X}$ to
derivatives with respect to $\mathbf{x}$ by assuming that
($P(\mathbf{x},k|\mathbf{X},t,Z)=P(\mathbf{x}+\mathbf{y},k|\mathbf{X}+\mathbf{y},t,Z)$
holds for all $\mathbf{y}$ (see assumption 2 in Section~\ref{TDSEb}).
Furthermore, it is understood that integrations are over the domain defined
by the measurement scenario.
Technically speaking, after passing to the continuum limit, $P(\mathbf{x}|\mathbf{X},t,Z)$
denotes the probability density, not the probability itself.
However, as mentioned above, we write integration and derivation symbols for both
the discrete case and its continuum limit and as there can be no confusion about
which case we are considering, we use the same symbol for
the probability density and the probability.

For later use, it is expedient to write Eq.~(\ref{LI0}) in a different form which
separates the data about the position of the clicks and the associated color $k$
as much as possible.
According to the product rule, we have
\begin{eqnarray}
P(\mathbf{x},k|\mathbf{X},t,Z)&=&P(k|\mathbf{x},\mathbf{X},t,Z)P(\mathbf{x}|\mathbf{X},t,Z)
,
\label{LI0a}
\end{eqnarray}
which we may, without loss of generality, represent as
\begin{eqnarray}
P(\mathbf{x},k=+1|\mathbf{X},t,Z)&=&P(\mathbf{x}|\mathbf{X},t,Z)\cos^2\frac{\theta(\mathbf{x},\mathbf{X},t,Z)}{2}
\nonumber \\
P(\mathbf{x},k=-1|\mathbf{X},t,Z)&=&P(\mathbf{x}|\mathbf{X},t,Z)\sin^2\frac{\theta(\mathbf{x},\mathbf{X},t,Z)}{2}
.
\label{LI1}
\end{eqnarray}

Substituting Eq.~(\ref{LI1}) into Eq.~(\ref{LI0}) we obtain

\begin{eqnarray}
I_F
&=& \int d\mathbf{x} \, dt
\left\{
\frac{1}{P(\mathbf{x}|\mathbf{X},t,Z)}\left[\bm\nabla P(\mathbf{x}|\mathbf{X},t,Z)\right]^2
+\left[\bm\nabla \theta(\mathbf{x},\mathbf{X},t,Z)\right]^2 P(\mathbf{x}|\mathbf{X},t,Z)
\right\}
,
\label{LI2}
\end{eqnarray}
which is the Fisher information for the measurement scenario described earlier.
Note that up to this point, we have not assumed that the particle moves or carries a magnetic moment nor
did we assign any particular meaning to $\theta(\mathbf{x},\mathbf{X},t,Z)$.

According to the principle laid out earlier, our task is to search
for the global minimum of Eq.~(\ref{LI2}), the Fisher information of the measurement scenario described above,
thereby excluding the uninformative class of solutions.

\subsection{Including knowledge}\label{two.e}
It is instructive to first search for the global minimum of
Eq.~(\ref{LI2}) in the case that we do not know whether the particle
moves or not and do not know about the effect of the applied electromagnetic field
on the frequency distribution of the $(\mathbf{j},k)_{n,\tau}$'s.
In this situation, we may discard the time dependence altogether
and search for the non-trivial global minimum of
\begin{eqnarray}
\widetilde I_F
&=& \int d\mathbf{x} \,
\left\{
\frac{1}{P(\mathbf{x}|\mathbf{X},Z)}\left[\bm\nabla P(\mathbf{x}|\mathbf{X},Z)\right]^2
+\left[\bm\nabla \theta(\mathbf{x},\mathbf{X},Z)\right]^2 P(\mathbf{x}|\mathbf{X},Z)
\right\}
.
\label{IK0}
\end{eqnarray}

For pedagogical purposes, we now specialize to the case of one spatial dimension
and discard the color dependence, that is we set $\bm\nabla \theta(\mathbf{x},\mathbf{X},Z)=0$
and assume that $\Omega\rightarrow [0,L]$ where $[0,L]$
is the range covered by the detection system.
With the latter assumption $P(x|X,Z)=0$ for $x\le0$ or $x\ge L$.

Recalling the assumption that space is homogeneous (see Eq.~(\ref{TDSE4})),
we search for solutions of the form $P(x|X,Z)=f(x-X,Z)$.
As $f(x-X,Z)\ge0$, we may substitute $P(x|X,Z)=f(x-X,Z)=\psi^2(x-X,Z)$
in Eq.~(\ref{IK0}) and we obtain
\begin{eqnarray}
\widetilde I_F
&=&
4\int_{0}^L dx \,
\left(\frac{\partial \psi(x-X,Z)}{\partial x}\right)^2
.
\label{IK1}
\end{eqnarray}
Recall that the requirement of a global minimum entails that $\widetilde I_F$ is
constant, independent of the unknown position $\mathbf{X}$ of the particle.

The extrema of Eq.~(\ref{IK1}) are easily found by a
standard variational calculation.
Introducing the Lagrange multiplier $\mu$ to account for the constraint
$\int_{0}^L dx \psi^2(x-X,Z)=\int_{0}^L dx  P(x|X,Z)=1$,
the extrema are the solutions of
\begin{eqnarray}
\frac{\partial^2 \psi(x-X,Z)}{\partial x^2}-\frac{\mu}{4} \psi(x-X,Z)&=&0
.
\label{IK2}
\end{eqnarray}
For $\mu>0$, the solutions of Eq.~(\ref{IK2}) are hyperbolic functions,
a family of solutions that is not compatible with the
constraint $P(x|X,Z)=0$ for $x=0,L$ and can therefore be ruled out.
Writing $\mu=-4\nu^2$, the general solution of Eq.~(\ref{IK2}) reads
\begin{eqnarray}
\psi(x-X,Z)&=&c_1(Z)\sin\nu (x-X) + c_2(Z)\cos\nu (x-X)
\nonumber \\
&=&[c_1(Z)\cos\nu X + c_2(Z)\sin\nu X ] \sin\nu x - [c_1(Z)\sin\nu X - c_2(Z)\cos\nu X ] \cos\nu x
,
\label{IK2a}
\end{eqnarray}
where $c_1(Z)$ and $c_2(Z)$ are integration constants.
Imposing the boundary condition $\psi(x-X,Z)=0$ for $x=0$
we must have $c_1(Z)\sin\nu X = c_2(Z)\cos\nu X$ hence the second
term in Eq.~(\ref{IK2a}) vanishes for all $x$.
In addition, imposing the boundary condition $\psi(x-X,Z)=0$ for $x=L$,
we must have either $c_1(Z)\cos\nu X + c_2(Z)\sin\nu X=0$
in which case $\psi(x-X,Z)=0$ for all $x$ or $\nu=n\pi/L$ for $n=1,2,\dots$
in which case the non-trivial solutions read
\begin{eqnarray}
\psi(x-X,Z)&=&\left[c_1(Z)\cos\frac{n\pi X}{L} + c_2(Z)\sin\frac{n\pi X}{L} \right] \sin\frac{n\pi x}{L} \quad,\quad n=1,2,\ldots
.
\label{IK2b}
\end{eqnarray}
Using $c_1(Z)\sin\nu X = c_2(Z)\cos\nu X$ with $\nu=n\pi/L$ we find that
\begin{eqnarray}
\psi^2(x-X,Z)&=&[c_1^2(Z) + c_2^2(Z)] \sin^2\frac{n\pi x}{L} \quad,\quad n=1,2,\ldots
,
\label{IK2c}
\end{eqnarray}
and from $\int_{0}^L dx\, \psi^2(x-X,Z)=1$ we find that $L[c_1^2(Z) + c_2^2(Z)]/2=1$.
Hence
\begin{eqnarray}
P(x|X,Z)&=&\frac{2}{L}\sin^2\frac{n\pi x}{L}\quad,\quad n=1,2,\ldots
,
\label{IK2d}
\end{eqnarray}
which are nothing but the solutions of the Schr\"odinger equation of a free particle
in a one-dimensional box~\cite{BALL03}.
Note that the r.h.s of Eq.~(\ref{IK2d}) does not depend on $X$.
In other words, from the measured data
we cannot infer anything about the unknown position $X$, in concert with the notion
that the particle is ``free''.
From Eq.~(\ref{IK2b}) it follows that $\widetilde I_F=(2n\pi/L)^2$, independent of $X$
as it should be. Clearly, the solution for non-trivial global minimum of $\widetilde I_F$
is given by Eq.~(\ref{IK2d}) with $n=1$.

Returning to the case that the frequency distribution of the $(\mathbf{j},k)_{n,\tau}$'s
indicates that the motion of the particle depends on the applied electric or magnetic field,
we can incorporate this additional knowledge as a constraint
on the global minimization problem.
In general, the global minimization problems that we will consider take the form
$\lambda I_F+\Lambda$ where $\lambda$ is a parameter (not a Lagrange multiplier) that ``weights'' the uncertainty
in the conditions (represented by $I_F$) relative to the knowledge represented by the functional
\begin{eqnarray}
\Lambda
&=& \int d\mathbf{x} \, dt
\sum_{k=\pm1}
F(\mathbf{x},k,t,Z) P(\mathbf{x},k|\mathbf{X},t,Z)
,
\label{IK3}
\end{eqnarray}
where $F(\mathbf{x},k,t,Z)$ is a function which encodes
the additional knowledge and which does not depend on the unknown position $\mathbf{X}$.

The assumption that space is homogeneous allows us to replace
derivatives with respect to $\mathbf{X}$ by derivatives with respect to $\mathbf{x}$.
This helps in searching for the global minimum of $\lambda I_F+ \Lambda$ because
it can be found by searching for the extrema of
\begin{eqnarray}
\lambda I_F+ \Lambda
&=& \int d\mathbf{x} \, dt
\sum_{k=\pm1}
\left\{
\frac{\lambda}{P(\mathbf{x},k|\mathbf{X},t,Z)}\left[\bm\nabla P(\mathbf{x},k|\mathbf{X},t,Z)\right]^2
+ F(\mathbf{x},k,t,Z) P(\mathbf{x},k|\mathbf{X},t,Z)
\right\}
,
\label{IK4}
\end{eqnarray}
as a functional of the $P(\mathbf{x},k|\mathbf{X},t,Z)$'s.
By the standard variational procedure, the extrema of $\lambda I_F+ \Lambda$
are the solutions of
\begin{eqnarray}
\frac{\lambda\left[\bm\nabla P(\mathbf{x},k|\mathbf{X},t,Z)\right]^2}{P^2(\mathbf{x},k|\mathbf{X},t,Z)}
+2\lambda\bm\nabla \left[\frac{\bm\nabla P(\mathbf{x},k|\mathbf{X},t,Z)}{P(\mathbf{x},k|\mathbf{X},t,Z)}\right]
- F(\mathbf{x},k,t,Z)&=&0
\quad,\quad k=1,2
,
\label{IK5}
\end{eqnarray}
On the other hand, the global minimum of $\lambda I_F+ \Lambda$ should not depend on unknown $\mathbf{X}$
because if it did, it was not a global minimum and in addition,
the values of $\lambda I_F+ \Lambda$ would tell us something about $\mathbf{X}$,
a contradiction to the assumption that $\mathbf{X}$ is unknown.

Taking the derivative of Eq.~(\ref{IK4}) with respect to $\mathbf{X}$ (recall $\mathbf{X}=\mathbf{X}_t$)
yields
\begin{eqnarray}
\bm\nabla_t
(\lambda I_F+ \Lambda)
&=& -\int d\mathbf{x} \, dt
\sum_{k=\pm1}
\left[
\lambda\frac{\left[\bm\nabla P(\mathbf{x},k|\mathbf{X},t,Z)\right]^2}{P^2(\mathbf{x},k|\mathbf{X},t,Z)}
+2\lambda\bm\nabla \left[\frac{\bm\nabla P(\mathbf{x},k|\mathbf{X},t,Z)}{P(\mathbf{x},k|\mathbf{X},t,Z)}\right]
-
 F(\mathbf{x},k,t,Z)
\right] \bm\nabla_t P(\mathbf{x},k|\mathbf{X},t,Z)
.
\label{IK6}
\end{eqnarray}
Comparing Eqs.~(\ref{IK5}) and~(\ref{IK6}) and recalling
the constraint  $\bm\nabla_\tau P(\mathbf{j},k|\mathbf{X}_{\tau},N,Z)\not=0$
used to eliminate uninformative solutions,
we conclude that the extrema (and therefore also the global minimum) of Eq.~(\ref{IK4}) are (is)
independent of $\mathbf{X}_t$, as required.

\subsection{Motion of the particle}\label{MP}\label{two.f}
We consider the limiting case that there is no uncertainty
on the position of the particle, that is $x=X$ for all clicks.
Then the motion of the particle and the motion of the positions of the detector clicks map one-to-one,
for each repetition of the experiment (by assumption).

From the data $\mathbf{x}(t)$ we can compute the vector field $\mathbf{U}(\mathbf{x},t)$ defined by
\begin{eqnarray}
\frac{d\mathbf{x}}{dt}&=& \mathbf{U}(\mathbf{x},t)
.
\label{MO0}
\end{eqnarray}
In principle, $\mathbf{U}(\mathbf{x},t)$ is
fully determined by the data obtained by repeating the experiment under different (initial) conditions.
In practice, however, it is unlikely that we have enough data to
compute $\mathbf{U}(\mathbf{x},t)$ for all $(\mathbf{x},t)$.

We only consider the case in which the position of the clicks
is encoded by its $(x,y,z)$-coordinates in an orthogonal frame of reference attached
to the observer.
Under the usual assumptions of differentiabilty etc.,
we can use the Helmholtz-like decomposition of a vector field
$\mathbf{U}(\mathbf{x},t)= \bm\nabla S(\mathbf{x},t) - \bm\nabla \times\mathbf{W}(\mathbf{x},t)$.
We will not use this form but write~\cite{RALS13b}
\begin{eqnarray}
\mathbf{U}(\mathbf{x},t)&=& \bm\nabla S(\mathbf{x},t)- \mathbf{A}(\mathbf{x},t)
,
\label{MO1}
\end{eqnarray}
where $S(\mathbf{x},t)$ is a scalar function and  $\mathbf{A}(\mathbf{x},t)$ a vector field.
Clearly Eq.~(\ref{MO1}) has some extra freedom which we can remove
by requiring that $ \mathbf{A}(\mathbf{x},t)=\bm\nabla \times\mathbf{W}(\mathbf{x},t)$.
This amounts to requiring that $\bm\nabla\cdot\mathbf{A}=0$.
It is convenient not do this at this stage so we take Eq.~(\ref{MO1})
and will impose $\bm\nabla\cdot\mathbf{A}=0$ later.
As mentioned earlier, if differentiabilty is an issue we
should use the finite-difference form of the $\bm\nabla$ operators.

For convenience, we drop the $(\mathbf{x},t)$ arguments and switch to a component-wise
notation in the few paragraphs that follow.
From Eq.~(\ref{MO0}) and Eq.~(\ref{MO1}) it directly follows that~\cite{RALS13b}
\begin{eqnarray}
\frac{d^2 x_i}{dt^2}&=&\frac{\partial U_i}{\partial t} +\sum_{j=1}^3 \frac{\partial U_i}{\partial x_j}U_j
\nonumber \\
&=&
\frac{\partial^2 S}{\partial x_i\partial t} -\frac{\partial A_i}{\partial t}+
\sum_{j=1}^3
\left(
\frac{\partial^2 S}{\partial x_i\partial x_j}-\frac{\partial A_i}{\partial x_j}
\right)
\left(
\frac{\partial S}{\partial x_j}-A_j
\right)
\nonumber \\
&=&
\frac{\partial}{\partial x_i}
\left[
\frac{\partial S}{\partial t}
+\frac{1}{2}\sum_{j=1}^3 \left(\frac{\partial S}{\partial x_j}-A_j\right)^2
\right]
+
\sum_{j=1}^3 \left(\frac{\partial A_j}{\partial x_i}-\frac{\partial A_i}{\partial x_j}\right)
\left(\frac{\partial S}{\partial x_j}-A_j\right)
-\frac{\partial A_i}{\partial t}
,
\label{MO2}
\end{eqnarray}
where $i=1,2,3$ labels the coordinate of the detector clicks.

Introducing the vector field $\mathbf{B}=\bm\nabla \times\mathbf{A}$
the second term in Eq.~(\ref{MO2}) can we written as
\begin{eqnarray}
\sum_{j=1}^3 \left(\frac{\partial A_j}{\partial x_i}-\frac{\partial A_i}{\partial x_j}\right)
\left(\frac{\partial S}{\partial x_j}-A_j\right)
&=&\left(\frac{d \mathbf{x}}{dt} \times \mathbf{B}\right)_i
.
\label{MO3}
\end{eqnarray}
It is important to note that in order to derive Eq.~(\ref{MO3}),
it is essential that the position is represented by three coordinates.
Switching back to the vector notation we have
\begin{eqnarray}
\frac{d^2 \mathbf{x}}{dt^2}&=&
\bm\nabla
\left[
\frac{\partial S}{\partial t}
+\frac{1}{2}\left(\bm\nabla S-\mathbf{A}\right)^2
\right]
+
\frac{d \mathbf{x}}{dt} \times \mathbf{B}
-\frac{\partial \mathbf{A}}{\partial t}
.
\label{MO4}
\end{eqnarray}

Up to now, we have not made any assumption other than that space is three-dimensional.
Next comes a crucial step in the reasoning.
Let us hypothesize that there exists a scalar field $\phi=\phi(\mathbf{x},t)$ such that
\begin{eqnarray}
\frac{\partial S}{\partial t}
+\frac{1}{2}\left(\bm\nabla S-\mathbf{A}\right)^2 &=& - \phi
.
\label{MO6}
\end{eqnarray}
Then, upon introducing the vector field $\mathbf{E}=-\bm\nabla \phi -\partial \mathbf{A}/\partial t$,
Eq.~(\ref{MO4}) becomes
\begin{eqnarray}
\frac{d^2 \mathbf{x}}{dt^2}&=&
\mathbf{E}+
\frac{d \mathbf{x}}{dt} \times \mathbf{B}
.
\label{MO7}
\end{eqnarray}
Although Eq.~(\ref{MO7}) has the same the structure as
the equation of motion of a charged particle in an electromagnetic field $(\mathbf{E},\mathbf{B})$,
our derivation of Eq.~(\ref{MO7}) is solely based on the elementary observation
that the data yields the vector field $\mathbf{U}(\mathbf{x},t)$ (see Eq.~(\ref{MO1})), some standard vector-field identities
and the hypothesis that there exist a scalar field $\phi$ such that Eq.~(\ref{MO6}) holds.
No reference to charged particles or electromagnetic fields enters the derivation.
Put differently (and putting aside technicalities related to differentiability),
if there exist a scalar field $\phi$ such that Eq.~(\ref{MO6}) holds, then
mathematics alone dictates that the equation of motion must have the structure Eq.~(\ref{MO7}),
with $\mathbf{E}$ and $\mathbf{B}$ having no relation to the electromagnetic field acting on a charged particle.
The latter relation is established when the data shows that there is indeed an effect of
electromagnetic field on the motion of the particle, an effect from which it is inferred that the particle carries charge.
This relation can be made explicit by introducing the symbols $m$ for the mass and $q$ for the charge
of the particle and by replacing $\mathbf{A}$ by $q\mathbf{A}/m$ (we work with MKS units throughout this paper)
and $\phi$ by $(q\phi + u)/m$ where $u$ represent all potentials that are not of electromagnetic origin.
Then we have

\begin{eqnarray}
m\frac{d^2 \mathbf{x}}{dt^2}&=&
-\bm\nabla u
+q\mathbf{E}+
{q}\frac{d \mathbf{x}}{dt} \times \mathbf{B}
,
\label{MO8}
\end{eqnarray}
and upon replacing $S$ by $S/m$ and $V=q\phi +u$
\begin{eqnarray}
\frac{\partial S}{\partial t}
+\frac{1}{2m}\left(\bm\nabla S-{q}\mathbf{A}\right)^2 + V &=&0
.
\label{MO9}
\end{eqnarray}
Note that we have obtained the Hamilton-Jacobi equation Eq.~(\ref{MO9})
without making any reference to a Hamiltonian, the action, contact transformations
and the like. In essence, Eqs.~(\ref{MO1})--(\ref{MO9}) follow from Eq.~(\ref{MO0}), some mathematical identities
and the crucial assumption that there exist a $V$ such that Eq.~(\ref{MO9}) holds.
Summarizing:
\begin{center}
\framebox{
\parbox[t]{0.9\hsize}{%
If we can find scalar fields $S$ and $V$ and a vector field
$\mathbf{A}(\mathbf{x},t)$ such that Eq.~(\ref{MO9}) holds for all $(\mathbf{x},t)$
then the clicks of the detectors will carve out a trajectory
that is completely determined
by the classical equation of motion Eq.~(\ref{MO8}) of a particle in a potential and subject to electromagnetic potentials.
}
}
\end{center}
\noindent
Of course, there is nothing really new in this statement: it is just telling us what we know from classical mechanics
but there is a slight twist.

First, given the data $\mathbf{x}(t)$ of the detector clicks, this data will not
comply with the equations of classical mechanics unless we can find
scalar fields $S$ (the action) and $V$ (the potential) and a vector field $\mathbf{A}(\mathbf{x},t)$
(vector potential) such that Eq.~(\ref{MO9}) holds.
Second, in the case of interest to us here, there is uncertainty on the mapping between the particle
position $\mathbf{X}(t)$ and the position of the corresponding clicks $\mathbf{x}(t)$
and there is no reason to expect that Eq.~(\ref{MO9}) will hold.
Instead of requiring that Eq.~(\ref{MO9}) holds, we will require that
there exists two scalar fields $V_{k}(\mathbf{x},t)$ for $k=\pm1$ such that
\begin{eqnarray}
\int d\mathbf{x} \, dt\,
\sum_{k=\pm1}
\left[
\frac{\partial S_k(\mathbf{x},t)}{\partial t}
+\frac{1}{2m}
\left(
\bm\nabla S_k(\mathbf{x},t)
-{q}\mathbf{A}(\mathbf{x},t)
\right)^2
+V_k(\mathbf{x},t)
\right]P(\mathbf{x},k|\mathbf{X},t,Z)
&=&0
,
\label{LI3}
\end{eqnarray}
where we regard the particles that respond with $k=+1$ or $k=-1$ as two different objects,
the clicks generated by each object being described by its own Hamilton-Jacobi equation with potentials $V_k(\mathbf{x},t)$.

The next step is to disentangle as much as possible
the motion of the positions of the clicks from their $k$-values.
We introduce $S_k(\mathbf{x},t)=S(\mathbf{x},t)-kR(\mathbf{x},t)$
for $k=\pm1$ and after some rearrangements we obtain
\begin{eqnarray}
\Lambda&=&\int d\mathbf{x} \, dt\,
\sum_{k=\pm1}
\left[
\frac{\partial S_k(\mathbf{x},t)}{\partial t}
+\frac{1}{2m}
\left(
\bm\nabla S_k(\mathbf{x},t)
-{q}\mathbf{A}(\mathbf{x},t)
\right)^2
+V_k(\mathbf{x},t)
\right]P(\mathbf{x},k|\mathbf{X},t,Z)
\nonumber \\
&=&\int d\mathbf{x} \, dt\,
\bigg\{
\frac{1}{2m}
\left[
\left(
\bm\nabla S(\mathbf{x},t)
-{q}\mathbf{A}(\mathbf{x},t)
\right)^2
+
\big(
\bm\nabla R(\mathbf{x},t)
\big)^2
-2\cos\theta(\mathbf{x},\mathbf{X},t,Z)
\bm\nabla R(\mathbf{x},t)
\left(
\bm\nabla S(\mathbf{x},t)
-{q}\mathbf{A}(\mathbf{x},t)
\right)
\right]
\nonumber \\
&&\hbox to 40pt{}
+
\left[
\frac{\partial S(\mathbf{x},t)}{\partial t}
-
\cos\theta(\mathbf{x},\mathbf{X},t,Z)\frac{\partial R(\mathbf{x},t)}{\partial t}
\right]
+V_0(\mathbf{x},t)+V_1(\mathbf{x},t)\cos\theta(\mathbf{x},\mathbf{X},t,Z)
\bigg\}
P(\mathbf{x}|\mathbf{X},t,Z)
,
\label{LI5}
\end{eqnarray}
where $V_0(\mathbf{x},t)=[V_{+1}(\mathbf{x},t)+V_{-1}(\mathbf{x},t)]/2$,
$V_1(\mathbf{x},t)=[V_{+1}(\mathbf{x},t)-V_{-1}(\mathbf{x},t)]/2$
and we made use of
$\sum_{k=\pm1} kP(\mathbf{x},k|\mathbf{X},t,Z)=\cos\theta(\mathbf{x},\mathbf{X},t,Z)P(\mathbf{x}|\mathbf{X},t,Z)$.
Omitting the terms involving $\cos\theta(\mathbf{x},\mathbf{X},t,Z)$ and $R(\mathbf{x},t)$,
Eq.~(\ref{LI5}) reduces to the
expression of the averaged Hamilton-Jacobi equation which entered
the derivation of the time-dependent Schr\"odinger equation~\cite{RAED14b}.

\subsection{Including the motion of the magnetic moment}\label{IM}

The function $\cos\theta(\mathbf{x},\mathbf{X},t,Z)$ determines
the ratio of $k=\pm1$ clicks
and $R(\mathbf{x},t)=(S_{-1}(\mathbf{x},t)-S_{+1}(\mathbf{x},t))/2$,
that is half of the difference between the actions of the $k=-1$ and $k=+1$ clicks.
We can relate these two functions to the direction
of a classical magnetic moment by imposing the constraint that
when the positions of the clicks (=particle position in this case) do not move, we recover the classical-mechanical equation
of motion of a magnetic moment in a magnetic field, for every $\mathbf{x}$.

In the limit that $m\rightarrow\infty$ (corresponding to the situation that the positions of the clicks hardly change with time)
we have
\begin{eqnarray}
\lim_{m\rightarrow\infty}\Lambda&=&\int d\mathbf{x} \, dt\,
\bigg\{
\left[
\frac{\partial S(\mathbf{x},t)}{\partial t}
-
\cos\theta(\mathbf{x},\mathbf{X},t,Z)\frac{\partial R(\mathbf{x},t)}{\partial t}
\right]
+V_0(\mathbf{x},t)+V_1(\mathbf{x},t)\cos\theta(\mathbf{x},\mathbf{X},t,Z)
\bigg\}
P(\mathbf{x}|\mathbf{X},t,Z)
.
\label{IM0}
\end{eqnarray}

Without loss of generality, we may assume that
$V_0(\mathbf{x},t)=\widetilde V_0(\mathbf{x},t)+\widehat V_0(\mathbf{x},t)$
where $\widetilde V_0(\mathbf{x},t)$
does not depend on $\theta(\mathbf{x},\mathbf{X},t,Z)$ and $R(\mathbf{x},t)$
while $\widehat V_0(\mathbf{x},t)$ may.
Writing $\widehat V_1(\mathbf{x},t)=\widehat V_0(\mathbf{x},t)+V_1(\mathbf{x},t)\cos\theta(\mathbf{x},\mathbf{X},t,Z)$,
searching for the extrema of Eq.~(\ref{IM0}) through
variation with respect to $\cos\theta(\mathbf{x},\mathbf{X},t,Z)$,
$R(\mathbf{x},t)$, $S(\mathbf{x},t)$ and $P(\mathbf{x},t)$ yields
\begin{eqnarray}
\frac{\partial R(\mathbf{x},t)}{\partial t}&=&
\frac{\partial \widehat V_1(\mathbf{x},t)}{\partial \cos\theta(\mathbf{x},\mathbf{X},t,Z) }
\label{IM1a}
\\
\frac{\partial \cos\theta(\mathbf{x},\mathbf{X},t,Z)}{\partial t}&=-&
\frac{\partial \widehat V_1(\mathbf{x},t)}{\partial R(\mathbf{x},t) }
\label{IM1b}
\\
S(\mathbf{x},t)\frac{\partial P(\mathbf{x}|\mathbf{X},t,Z)}{\partial t}&=&0
\label{IM1c}
\\
\frac{\partial S(\mathbf{x},t)}{\partial t}
&=&
\cos\theta(\mathbf{x},\mathbf{X},t,Z)\frac{\partial R(\mathbf{x},t)}{\partial t}
-\widehat V_0(\mathbf{x},t)-\widehat V_1(\mathbf{x},t)
.
\label{IM1d}
\end{eqnarray}
From Eq.~(\ref{IM1c}) it follows that $P(\mathbf{x}|\mathbf{X},t,Z)$ does not change
with time, in concert with the assumption that the positions of the clicks are stationary.
Comparing Eqs.~(\ref{IM1a}) and~(\ref{IM1b}) with Eq.~(\ref{MM5}), it is clear
that we will recover the classical equations of motion of the magnetic moment
if (i) we set $\widehat V_1(\mathbf{x},t)=-\gamma \mathbf{m}(\mathbf{x},t)\cdot\mathbf{B}(\mathbf{x},t)$
where $\mathbf{m}(\mathbf{x},t)$ is a unit vector, and
(ii) make the symbolic identification $z=\cos\theta(\mathbf{x},\mathbf{X},t,Z)$
and $\varphi(\mathbf{x},t)=R(\mathbf{x},t)/\unitS$ where $\unitS$ needs to be introduced to give
$\unitS\varphi(\mathbf{x},t)$ the dimension of $S(\mathbf{x},t)$.
Substituting the infered expression for $\widehat V_1(\mathbf{x},t)$ in Eq.~(\ref{LI5}) yields
\begin{eqnarray}
\Lambda
&=&\int d\mathbf{x} \, dt\,
\bigg\{
\frac{1}{2m}
\left[
\left(
\bm\nabla S(\mathbf{x},t)
-{q}\mathbf{A}(\mathbf{x},t)
\right)^2
+
\unitS^2
\big(
\bm\nabla \varphi(\mathbf{x},t)
\big)^2
-2\unitS\cos\theta(\mathbf{x},\mathbf{X},t,Z)
\bm\nabla \varphi(\mathbf{x},t)
\left(
\bm\nabla S(\mathbf{x},t)
-{q}\mathbf{A}(\mathbf{x},t)
\right)
\right]
\nonumber \\
&&\hbox to 40pt{}
+
\left[
\frac{\partial S(\mathbf{x},t)}{\partial t}
-
\unitS
\cos\theta(\mathbf{x},\mathbf{X},t,Z)\frac{\partial \varphi(\mathbf{x},t)}{\partial t}
\right]
+V_0(\mathbf{x},t)
-a\gamma \mathbf{m}(\mathbf{x},t)\cdot\mathbf{B}(\mathbf{x},t)
\bigg\}
P(\mathbf{x}|\mathbf{X},t,Z)
.
\label{LI6}
\end{eqnarray}

\end{widetext}

\subsection{Derivation of the Pauli equation}\label{DPE}

We now have all ingredients to derive the Pauli equation from the principle
that logical inference applied to the most robust experiment yields a quantum theoretical description~\cite{RAED14b}.
According to this principle, we should search for the global minimum
of the Fisher information for the experiment, subject to the condition
that when the uncertainty vanishes, we recover the equations of motion of classical mechanics~\cite{RAED14b}.
Thus, we should search for the global minimum of
\begin{eqnarray}
F=\lambda I_F+ \Lambda
,
\label{DPE0}
\end{eqnarray}
where $I_F$ and $\Lambda$ are given by Eqs.~(\ref{LI2}) and~(\ref{LI6}), respectively.

In Appendix~\ref{sec1}, it is shown that the quadratic functional $Q$ which yields the Pauli equation
is identical to Eq.~(\ref{DPE0}) if we
make the identification $V_0(\mathbf{x},t)=q\phi(\mathbf{x},t)$,
$\unitS=\hbar/2$, $\gamma=q/m$ and $\lambda=\hbar^2/8m$ and
\begin{eqnarray}
\Phi(\mathbf{x},t)
&=&
\left(\begin{array}{r}
P^{1/2}(\mathbf{x},k=+1|\mathbf{X},t,Z)e^{iS_1(\mathbf{x},t)/\hbar}\\
P^{1/2}(\mathbf{x},k=-1|\mathbf{X},t,Z)e^{iS_2(\mathbf{x},t)/\hbar}
\end{array}\right)
.
\label{DPE1}
\end{eqnarray}

\begin{center}
\framebox{
\parbox[t]{0.9\hsize}{%
This then completes the derivation of the Pauli equation from logical inference principles.
}
}
\end{center}

\subsection{Discussion}

In Section~\ref{two.f}, we showed how to include the knowledge that in the absence of uncertainty the particle's motion is described by
Newtonian mechanics. Obviously, this treatment requires the particle to have a nonzero mass. On the other hand, in our logical
inference treatment of the free particle in Section II.E, the notion of mass does not enter in the derivation of Eq.~(\ref{IK2d}) but
neither does the concept of motion. This raises the interesting question how to inject into the logical inference treatment the
notion of moving massless particles with spin. We believe that the analogy with the pseudo-spin in graphene mentioned in the
introduction may provide a fruitful route to explore this issue.

The carbon atoms of ideal single-layer graphene form a
hexagonal lattice with the $\pi$-band (originating from $p_z$-orbitals of carbon atoms) well separated from other bands~\cite{KATS12}.
The electronic band structure of graphene has the remarkable feature that in the continuum limit, the low-energy excitations are
described by the two-dimensional Dirac equation for two species of massless fermions (corresponding to two valleys, $K$ and $K'$).
The fact that there the wave function of each of these two species is a two-component ``spinor'' is not related to the intrinsic
spin of the electron but is a manifestation of the two sub-lattice and bipartite structure of the hexagonal
lattice~\cite{KATS12}. This feature (Dirac-like spectrum) is present already in the simplest model where only the
nearest-neighbor hopping is taken into account~\cite{WALL47} but, actually, it is robust and follows just from discrete
symmetries, namely, time-reversal and inversion symmetries~\cite{KATS12}. A generalization to a four dimensional lattice,
retaining the property that the continuum limit yields the Dirac equation, is given in Ref.~\onlinecite{CREU08}. This is a nice
illustration of the fact that the model of a rotating electron is not the only way to arrive at the concept of spin. In our
derivation of the Pauli equation, we have to make the additional assumption (based on experimental observations such as the
anomalous Zeeman effect) that the interaction of this intrinsic degree of freedom with an external magnetic field is described
by the standard classical expression for the energy of a magnetic moment.

The next important step might be the derivation of
the Dirac equation. The Creutz model~\cite{CREU08} suggests that we should consider incorporating into the logical inference
treatment, the additional knowledge that one has objects hopping on a lattice instead of particles moving in a space-time
continuum.  Recall that up to Section II.D, the description of the measurement scenario, robustness etc. is explicitly discrete.
In Section~\ref{two.d}, the continuum limit was taken only because our aim was to derive the Pauli equation, which is formulated in
continuum space-time.  Of course, the description of the motion of the particle in Section~\ref{two.f} is entirely within a continuum
description but there is no fundamental obstacle to replace this treatment by a proper treatment of objects hopping on a
lattice. Therefore it seems plausible that the logical inference approach can be extended to describe massless spin-1/2
particles moving in continuum space-time by considering the continuum limit of the corresponding lattice model.
An in-depth, general treatment of this problem is beyond the scope of the present
paper and we therefore leave this interesting problem for future research.

A comment on the appearance of $\hbar$ is in order.
First of all, it should be noted that recent work has shown that $\hbar$
may be eliminated from the basic equations of (low-energy) physics
by a re-definition of the units of mass, time, etc.~\cite{VOLO10,RALS13a}.
This is also clear from the way $\hbar$ appears in the identification
that we used to shown that
quadratic functional $Q$ which yields the Pauli equation (see Eq.~(\ref{PA0}))
is the same as Eq.~(\ref{DPE0}).
With the MKS units adopted in the present paper,
Planck's constant $\hbar$ enters because of dimensional reasons ($a=\hbar/2$)
and also controls the importance of the term that expresses
the robustness of the experimental procedure ($\lambda=\hbar^2/8m$).
The actual value of $\lambda$ can only be determined by laboratory experiments.
Note that the logical-inference derivation of the canonical ensemble of statistical mechanics~\cite{JAYN57a,JAYN57b}
employs the same reasoning to relate the inverse temperature $\beta=1/k_BT$ to the average thermal energy.

We end this section by addressing a technicality.
Mappings such as Eq.~(\ref{DPE1}) are not one-to-one.
This is clear: we can alway add a multiple of $2\pi\hbar$
to $S_1(\mathbf{x},t)$ or $S_2(\mathbf{x},t)$, for instance.
In the hydrodynamic form of the Schr\"odinger equation~\cite{MADE27},
the ambiguity that ensues has implications for the interpretation
of the gradient of action as a velocity field~\cite{WALL94,VOLO09}.
As pointed out by Novikov, similar ambiguities appear in classical mechanics proper
if the local equations of motion (Hamilton equations) are not sufficient to characterize
the system completely and the global structure of the phase space
has to be taken into consideration~\cite{NOVI82}.
However, for the present purpose, this ambiguity has no effect on the minimization
of $F$ because Eq.~(\ref{DPE0}) does not change
if we add to $S_1(\mathbf{x},t)$ or $S_2(\mathbf{x},t)$ a real number which does not depend on $(\mathbf{x},t)$
(as is evident from Eq.~(\ref{LI5})) or,
equivalently, if we multiply $\Phi(x|\mathbf{X},t,Z)$ by a global phase factor
and add a constant to $\varphi(\mathbf{x},t)$.

\section{Stern-Gerlach experiment: neutral magnetic particle}\label{SGE}

The Stern-Gerlach experiment with silver atoms~\cite{STER22} and neutrons~\cite{SHER54}
demonstrates that a magnetic field affects the motion of a neutral particle
suggesting that minimalist theoretical description should account
for the interaction of the magnetic moment of the particle
and the applied magnetic field.
As is clear from the definition of the Pauli Hamiltonian
Eq.~(\ref{PA1}), in the Pauli equation
the magnetic field is directly linked to the charge $q$ of the particle.
Therefore, in this form the Pauli equation cannot be used to
describe the motion of a neutral magnetic particle in a magnetic field.

In quantum theory, this problem is solved by the ad-hoc
introduction of the intrinsic magnetic moment which is proportional to the spin
and by replacing $q\hbar/2m$ by the gyromagnetic ratio $\gamma$,
the value of which is particle-specific.

In the logical-inference treatment, no such ad-hoc procedure is necessary.
We simply set $q=0$ in Eq.~(\ref{LI6}) and use Eq.~(\ref{DPE1})
to find the equivalent quadratic form.
The Hamiltonian that appears in this quadratic form reads
\begin{eqnarray}
H&=&
-\frac{\hbar^2}{2m}\bm\nabla^2-\gamma\bm{\sigma}\cdot\mathbf{B}(\mathbf{x},t)
,
\label{SGE0}
\end{eqnarray}
where $\gamma$ is the gyromagnetic ratio which, in general, is not given by $q/m$.
As mentioned earlier, the appearance in Eq.~(\ref{SGE0}) of the Pauli-matrices
is a direct consequence of logical inference applied to
robust experiments that yield data in the form
of the position and one of the two kinds of detector clicks.

\section{Relation to earlier work}\label{EAR}

Readers familiar with the hydrodynamic formulation of quantum theory~\cite{MADE27}
and its interpretation in terms of Bohmian mechanics~\cite{BOHM52,TAKA52}
undoubtedly recognize the steps which transform quadratic functional Eq.~(\ref{PA0}) yielding the
Pauli equation Eq.~(\ref{PA1}) and the functional $Q$ given by Eq.~(\ref{PA14}).
In fact, the functional $Q$ has been used as the starting point
for the hydrodynamic representation~\cite{TAKA54}
and a causal interpretation~\cite{BOHM55a,BOHM55b,SCHI62} of the Pauli equation.
In this formulation, the two-component spinor can be given a classical-mechanics
interpretation in terms of an assembly of very small rotating bodies
which are distributed continuously in space.
Within this interpretations spins of different bodies interact.

Clearly, the logical-inference treatment does not support this interpretation:
the functional Eq.~(\ref{PA14}) is the result of analyzing
a robust experiment that yields data in the form of $(\mathbf{x},k)$ where
$\mathbf{x}$ is a 3-dimensional coordinate and $k=\pm1$ denotes the two-valued ``color'',
together with the requirement that on average and in special cases,
the data should comply with the classical-mechanical motion.

An expression of Eq.~(\ref{PA14}) in which the separation of the
contribution of the Fisher information and the classical-field mechanical
is explicit has been given by Reginatto~\cite{REGI98b}.
This expression is different from ours.
Comparing  Eq.~(\ref{LI2}) with Eq.~(6,7) in Ref.\onlinecite{REGI98b},
we find that the expressions are fundamentally different
due the fact that the representation (7), when substituted
in (6), does not yield Eq.~(\ref{PA14}).

\section{Conclusion}\label{CON}

It is somewhat discomforting that it takes a considerable amount of symbolic manipulations
to derive the Pauli equation from the combination of the measurement scenario,
the notion of a robust experiment and the behavior
expected in some limiting cases.
Therefore, it may be worthwhile to recapitulate what has be done in simple words,
without worrying too much about the technicalities.

The first step is to describe the measurement scenario.
It is assumed that the object (particle) we are interested in
responds to the signal that we send to probe it.
The response of the object triggers a detection event.
In the case at hand, the data representing the detector clicks
consist of spatial coordinates and two-valued ``color'' indices.
We assign an i-prob to the whole data set.
To make progress, it is necessary to make assumptions about
the data-collection procedure.
We assume that each time we probe the object,
the data produced by the detection system is
logically independent of all other data produced
by previous/subsequent probing.
With this assumption, together with the assumption
that is does not matter where we carry out the experiment,
the notion of a robust experiment is found
to be equivalent to the global minimum of the Fisher information
for the corresponding measurement scenario (see Eq.~(\ref{LI2})).

The next step is to bring in the knowledge that in the
extreme case that there is no uncertainty about the
outcome of each detection event, we expect to observe
data that is compliant with classical, Newtonian mechanics
both for the motion of a particle
as well as for the motion of its magnetic moment
in the case that the particle does not move (see Eq.~(\ref{LI6})).

The third step is to find the balance between the uncertainty
in the detection events represented by Eq.~(\ref{LI2})
and the ``classical mechanics'' knowledge represented by Eq.~(\ref{LI6})
by searching for the global minimum of Eq.~(\ref{DPE0}) for
all possible unknown positions of the particle.
The result of this calculation is a fairly complicated
non-linear set of equations for the i-prob to observe a click.

The final step is to observe that by transformation Eq.~(\ref{DPE1}),
this non-linear set of equations and the Pauli equation are
equivalent.
The latter, being a set of linear equations, are much easier
to solve than their non-linear equivalent.

In the logical inference approach, the assumption that
each time we probe the object, the detection system
reports a two-valued ``color'' index
and our requirement that in the extreme case mentioned earlier
we expect to see the motion of a classical magnetic moment
automatically leads to the notion of a ``quantized''
(i.e. two-valued) intrinsic magnetic moment.
The notion of spin appears as an inference, forced upon us
by the (two-valued) data and our assumptions
(which do not make reference to concepts of quantum theory)
that the experiment is robust, etc.

From a more general perspective, it is remarkable
that the logic inference approach introduces the concept of ``spin''
in a way which is not much different from the way real numbers are introduced.
Indeed, the latter appear as a necessity to provide an answer to questions such as
``what new kind of number do we have to introduce such that
the square of it yields the integer $n$''.
If $n=m^2$ where $m$ is an integer, no new concept has to be introduced
but if say $n=2$, the answer to the question is given the symbolic name $\sqrt{2}$.

Similarly, in our logical-inference treatment the concept of spin
naturally appears as a result of describing situations in which there is two-valued data
and the requirement that in a limiting case we recover the classical equation of motion.
This concept of spin only exists in our mind, in complete agreement with the fact
that this concept maybe put to very good use whenever there are two-valued variables that
may or may not relate to (intrinsic) angular momentum,
as in the theory of the electronic properties of graphene, for example~\cite{KATS12}.

It will not have escaped the reader that
in the logical-inference derivation of the Pauli equation as well as in earlier work along this line~\cite{RAED14b,RAED13b}
there are no postulates regarding ``wavefunctions'', ``observables'', ``quantization rules',
no ``quantum'' measurements~\cite{NIEU13},``Born's rule'', etc.
This is a direct consequency of the basic premise of this approach, namely that
current scientific knowledge derives, through cognitive processes in the human brain, from the discrete events which
are observed in laboratory experiments and from relations between those events that we, humans, discover.
These discrete events are not ``generated'' according to certain quantum laws: instead these laws
appear as the result of (the best) inference based on available data in the form of discrete events.
In essence, for all the basic but fundamental cases treated so far,
the machinery of quantum theory appears as a result of transforming a set of non-linear equations
into a set on linear ones.
The wavefunction, spinor, spin, \ldots are
all mathematical concepts, vehicles that render a class of complicated nonlinear minimization problems
into the minimization of a quadratic forms.
As products of our collective imagination,
these concepts are extraordinarily useful but have no tangible existence, just like numbers themselves.
Of course, it remains to be seen whether the logical-inference approach
can be extended to e.g. many-body and relativistic quantum physics.

In summary:
the Pauli equation derives from logical inference applied
to robust experiments in which there is uncertainty about individual detection events
which yield information about the particle position and its two-valued ``color''.
This derivation adds another, new instance to the list of examples~\cite{RAED14b,RAED13b}
for which the logical-inference approach establishes a bridge between objective knowledge gathered
through experiments and their description in terms of concepts.

\section*{Acknowledgement}
We would like to thank Koen De Raedt, Karl Hess,
Thomas Lippert, and Seiji Miyashita for many stimulating discussions.
MIK and HCD acknowedges a financial support by European Research
Council, project 338957 FEMTO/NANO.

\appendix
\section{The algebra of logical inference}\label{sec2}

This appendix does not contain any original material
but is provided to render the present paper self-contained.

If we are only concerned about quantifying the truth of a proposition
given the truth of another proposition,
it is possible to construct a mathematical framework, an extension of Boolean logic,
that allows us to reason in a manner which is unambiguous and independent of the
individual, in particular if there are elements of uncertainty in the description~\cite{COX46,COX61,TRIB69,SMIT89,JAYN03}.

The algebra of logical inference can be derived~\cite{COX61,TRIB69,SMIT89,JAYN03}
from three so-called ``desiderata''.
The formulation which follows is taken from Ref.~\onlinecite{SMIT89}.

\noindent
{\bf Desideratum 1.}
{\it Plausibilities are represented by real numbers.}
The plausibility that a proposition $A$
is true conditional on proposition $B$ being true
will be denoted by $P(A|B)$.

\noindent
{\bf Desideratum 2.}
{\it Plausibilities must exhibit agreement with rationality.}
As more and more evidence supporting
the truth of a proposition becomes available,
the plausibility should increase monotonically and continuously
and the plausibility of the negation of the proposition
should decrease monotonically and continuously.
Moreover, in the limiting case that proposition $A$ is known to be either true
or false, the plausibility $P(A|B)$ should conform to the rules of deductive reasoning.
In other words, plausibilities must be in qualitative agreement with
the patterns of plausible reasoning uncovered by P\'olya~\cite{POLY54}.

\noindent
{\bf Desideratum 3.}
{\it All rules relating plausibilities must be consistent.}
Consistency of rational reasoning demands that if the rules
of logical inference allow a plausibility to be obtained in more than one way,
the result should not depend on the particular sequence of operations.
These three desiderata only describe the essential features of the plausibilities
and definitely do not constitute a set of axioms which plausibilities have to satisfy.

It is a most remarkable fact that these three desiderata suffice to uniquely determine
the set of rules by which plausibilities may be manipulated~\cite{COX61,TRIB69,SMIT89,JAYN03}.
Omitting the derivation, it follows that plausibilities
may be chosen to take numerical values in the range $[0,1]$ and
obey the rules~\cite{COX61,TRIB69,SMIT89,JAYN03}
\begin{enumerate}
\item
$P(A|Z)+P({\bar A}|Z)=1$ where
${\bar A}$ denotes the negation of proposition $A$
and $Z$ is a proposition assumed to be true.
\item
$P(AB|Z)=P(A|BZ)P(B|Z)=P(B|AZ)P(A|Z)$ where
the ``product'' $BZ$ denotes the logical product (conjunction)
of the propositions $B$ and $Z$,
that is the proposition $BZ$ is true if both $B$ and $Z$ are true.
This rule will be referred to as ``product rule''.
It should be mentioned here that it is not allowed
to define a plausibility for a proposition
conditional on the conjunction of mutual exclusive
propositions. Reasoning on the basis of two or more contradictory premises
is out of the scope of the present paper.
\item
$P(A{\bar A}|Z)=0$ and $P(A+{\bar A}|Z)=1$
where the ``sum'' $A+B$ denotes the logical sum (inclusive disjunction)
of the propositions $A$ and $B$,
that is the proposition $A+B$ is true if either $A$ or $B$ or both are true.
These two rules show that
Boolean algebra is contained in the algebra of plausibilities.
\end{enumerate}

The algebra of logical inference, as defined by the rules (1--3),
is the foundation for powerful tools such as the maximum entropy method
and Bayesian analysis~\cite{TRIB69,JAYN03}.
The rules (1--3) are unique~\cite{TRIB69,SMIT89,JAYN03}.
Any other rule which applies to plausibilities represented by real numbers
and is in conflict with rules (1--3)
will be at odds with rational reasoning and consistency,
as embodied by the desiderata 1--3.

The rules (1--3) are identical
to the rules by which we manipulate probabilities~\cite{KEYN21,FELL68,GRIM95,JAYN03}.
However, the rules (1--3) were not postulated.
They were derived from general considerations about
rational reasoning and consistency only.
Moreover, concepts such as sample spaces,
probability measures etc., which are an essential part of the mathematical
foundation of probability theory~\cite{FELL68,GRIM95}, play no role
in the derivation of rules (1--3).
Perhaps most important in the context of quantum theory
is that in the logical inference approach uncertainty about an event does not
imply that this event can be represented by a random variable
as defined in probability theory~\cite{GRIM95}.

There is a significant conceptual difference
between ``mathematical probabilities'' and plausibilities.
Mathematical probabilities are elements of an axiomatic framework
which complies with the algebra of logical inference.
Plausibilities are elements of a language
which also complies with the algebra of logical inference
and serve to facilitate communication,
in an unambiguous and consistent manner,
about phenomena in which there is uncertainty.

The plausibility $P(A|B)$ is an intermediate mental construct that
serves to carry out inductive logic, that is rational reasoning,
in a mathematically well-defined manner~\cite{TRIB69}.
In general, $P(A|B)$ may express the degree of believe of an individual that
proposition $A$ is true, given that proposition $B$ is true.
However, in the present paper, we explicitly exclude applications of this
kind because they do not comply with our main goal,
namely to describe phenomena ``in a manner independent
of individual subjective judgment''.

\begin{center}
\framebox{
\parbox[t]{0.9\hsize}{%
To take away this subjective connotation of the word ``plausibility'',
we will simply call $P(A|B)$ the ``inference-probability'' or ``i-prob'' for short.
}
}
\end{center}

A comment on the notation used throughout this paper is in order.
To simplify the presentation, we make no distinction between an event such as ``detector D has fired''
and the corresponding proposition ``$D =$ detector D has fired''.
If we have two detectors, say $D_x$ where $x=\pm1$, we write
$P(x|Z)$ to denote the i-prob
of the proposition
that detector $D_x$
fires, given that proposition $Z$ is true.
Similarly, the i-prob of the proposition
that two detectors $D_x$ and $D_y$ fire, given that proposition $Z$ is true,
is denoted by $P(x,y|Z)$.
Obviously, this notation generalizes to more than two propositions.

\section{Pauli equation: quantum theory}\label{sec1}

In this appendix, we show that the quadratic form, the minimization of
which yields the Pauli equation, is identical to the one
derived in Section~\ref{LIP} through logical inference.

The Pauli equation for a particle with mass $m$ and
charge $q$ can be written as

\begin{eqnarray}
i\hbar\frac{\partial}{\partial t}\Phi &=& H \Phi
,
\label{TDSE0}
\end{eqnarray}
where
\begin{eqnarray}
\Phi&=&\Phi(\mathbf{x},t)=
\left(\begin{array}{r}
        \Phi_1(\mathbf{x},t)\\
        \Phi_2(\mathbf{x},t)
\end{array}\right)
,
\label{PA1}
\end{eqnarray}
is a two-component wavefunction and the Hamiltonian is given by
\begin{eqnarray}
H&=&\frac{1}{2m}\left\{ \bm{\sigma}\cdot \left[-i\hbar\bm\nabla-{q}\mathbf{A}(\mathbf{x},t)\right]\right\}^2+q\phi(\mathbf{x},t)
\nonumber\\
&=&
\frac{1}{2m}\left[-i\hbar\bm\nabla-{q}\mathbf{A}(\mathbf{x},t)\right]^2
+q\phi(\mathbf{x},t)-\frac{q\hbar}{2m}\bm{\sigma}\cdot\mathbf{B}(\mathbf{x},t)
,
\nonumber \\
\label{PA1a}
\end{eqnarray}
where $\bm{\sigma}=(\sigma^x,\sigma^y,\sigma^z)^T$ denote the Pauli-spin matrices.
\begin{widetext}

By the standard variational argument, it follows that the Pauli equation is an extremum of the
quadratic form (functional)
\begin{eqnarray}
Q= \int d\mathbf{x} \, dt
&\bigg[&
\frac{i\hbar}{2}
\bigg(
\frac{\partial \Phi^\dagger }{\partial t}\Phi
-\Phi^\dagger \frac{\partial \Phi}{\partial t}
\bigg)
+\frac{1}{2m}
\bigg(+i\hbar\bm\nabla\Phi^\dagger -{q}\mathbf{A}(\mathbf{x},t)\Phi^\dagger\bigg)
\bigg(-i\hbar\bm\nabla\Phi -{q}\mathbf{A}(\mathbf{x},t)\Phi\bigg)
\nonumber \\&&
+q\phi(\mathbf{x},t)\Phi^\dagger\Phi
-\frac{q \hbar}{2m}\Phi^\dagger\bm{\sigma}\cdot \mathbf{B}(\mathbf{x},t)\Phi
\bigg]
,
\label{PA0}
\end{eqnarray}
with respect to variations in $\Phi^\dagger$.
We want to know how Eq.~(\ref{PA0}) looks like when we substitute
the polar representation
\begin{eqnarray}
\Phi
&=&
\left(\begin{array}{r}
\sqrt{P_1(\mathbf{x},t)}e^{iS_1(\mathbf{x},t)/\hbar}\\
\sqrt{P_2(\mathbf{x},t)}e^{iS_2(\mathbf{x},t)/\hbar}
\end{array}\right)
,
\label{PA2}
\end{eqnarray}
for both components of the spinor.
We have
\begin{eqnarray}
\frac{\partial \Phi_k(\mathbf{x},t)}{\partial t}
&=&
\left[
\frac{1}{2P_k(\mathbf{x},t)}\frac{\partial P_k(\mathbf{x},t)}{\partial t}
+\frac{i}{\hbar}
\frac{\partial S_k(\mathbf{x},t)}{\partial t}
\right]
\sqrt{P_k(\mathbf{x},t)}e^{iS_k(\mathbf{x},t)/\hbar}
,
\label{PA3}
\end{eqnarray}
for $k=1,2$ and
\begin{eqnarray}
\Phi^\dagger
\frac{\partial \Phi}{\partial t}
&=&
\sum_{k=1}^2
\left[
\frac{1}{2P_k(\mathbf{x},t)}\frac{\partial P_k(\mathbf{x},t)}{\partial t}
+\frac{i}{\hbar}
\frac{\partial S_k(\mathbf{x},t)}{\partial t}
\right]
P_k
,
\label{PA4}
\end{eqnarray}
from which it directly follows that
\begin{eqnarray}
\frac{i\hbar}{2}
\left(
\frac{\partial \Phi^\dagger }{\partial t}\Phi
-
\Phi^\dagger \frac{\partial \Phi}{\partial t}
\right)
&=&
\frac{\partial S_1(\mathbf{x},t)}{\partial t}
P_1(\mathbf{x},t)
+
\frac{\partial S_2(\mathbf{x},t)}{\partial t}
P_2(\mathbf{x},t)
.
\label{PA5}
\end{eqnarray}
Likewise we have
\begin{eqnarray}
-i\hbar
\left(\bm\nabla\Phi_k
-\frac{iq}{\hbar}\mathbf{A}(\mathbf{x},t)\Phi_k\right)
&=&
-i\hbar
\left[
\frac{1}{2P_k(\mathbf{x},t)}\bm\nabla P_k(\mathbf{x},t)
+\frac{i}{\hbar}\bm\nabla S_k(\mathbf{x},t)
-\frac{iq}{\hbar}\mathbf{A}(\mathbf{x},t)
\right]
\sqrt{P_k(\mathbf{x},t)}e^{iS_k(\mathbf{x},t)/\hbar}
,
\label{PA6}
\end{eqnarray}
and
\begin{eqnarray}
\frac{\hbar^2}{2m}
\left(\bm\nabla\Phi^\dagger +\frac{iq}{\hbar}\mathbf{A}(\mathbf{x},t)\Phi^\dagger\right)
&&\left(\bm\nabla\Phi -\frac{iq}{\hbar}\mathbf{A}(\mathbf{x},t)\Phi\right)
=
\nonumber \\
&&
\frac{\hbar^2}{2m}
\sum_{k=1}^2
\left\{
\frac{1}{4P_k^2(\mathbf{x},t)}\bigr[\bm\nabla P_k(\mathbf{x},t)\bigr]^2
+\frac{1}{\hbar^2}
\bigr[
\bm\nabla S_k(\mathbf{x},t) -{q}\mathbf{A}(\mathbf{x},t)
\bigr]^2
\right\}
P_k(\mathbf{x},t)
.
\label{PA7}
\end{eqnarray}
Furthermore, it follows that
\begin{eqnarray}
\phi(\mathbf{x},t)\Phi^\dagger\Phi&=&\phi(\mathbf{x},t)\left[P_1(\mathbf{x},t)+P_2(\mathbf{x},t)\right]
,
\\
\Phi^\dagger\sigma^x \Phi&=&2\sqrt{P_1(\mathbf{x},t)P_2(\mathbf{x},t)}\cos\frac{S_2(\mathbf{x},t)-S_1(\mathbf{x},t)}{\hbar},
\\
\Phi^\dagger\sigma^y \Phi&=&2\sqrt{P_1(\mathbf{x},t)P_2(\mathbf{x},t)}\sin\frac{S_2(\mathbf{x},t)-S_1(\mathbf{x},t)}{\hbar},
\\
\Phi^\dagger\sigma^z \Phi&=&P_1(\mathbf{x},t)-P_2(\mathbf{x},t)
.
\label{PA8}
\end{eqnarray}
Thus, we have all the expressions to write Eq.~(\ref{PA0}) in terms of
$P_1(\mathbf{x},t)$,
$P_2(\mathbf{x},t)$,
$S_1(\mathbf{x},t)$, and
$S_2(\mathbf{x},t)$.

Without loss of generality and without assigning a particular meaning to the new symbols yet,
we write
\begin{eqnarray}
P_1(\mathbf{x},t)&=&P(\mathbf{x},t)\cos^2\frac{\theta(\mathbf{x},t)}{2}
\nonumber \\
P_2(\mathbf{x},t)&=&P(\mathbf{x},t)\sin^2\frac{\theta(\mathbf{x},t)}{2}
.
\label{PA9}
\end{eqnarray}
Then we have
\begin{eqnarray}
\sum_{k=1}^2
\frac{1}{4P_k(\mathbf{x},t)}\left(\bm\nabla P_k(\mathbf{x},t)\right)^2
&=&
\frac{1}{4}
\left\{
\frac{1}{P_1(\mathbf{x},t)}\left[
\bm\nabla P(\mathbf{x},t)\cos^2\frac{\theta(\mathbf{x},t)}{2}
-
P(\mathbf{x},t)\cos\frac{\theta(\mathbf{x},t)}{2}\sin\frac{\theta(\mathbf{x},t)}{2} \bm\nabla \theta(\mathbf{x},t)
\right]^2
\right.
\nonumber \\
&&+
\left.
\frac{1}{P_2(\mathbf{x},t)}\left[
\bm\nabla P(\mathbf{x},t)\sin^2\frac{\theta(\mathbf{x},t)}{2}
+
P(\mathbf{x},t)\cos\frac{\theta(\mathbf{x},t)}{2}\sin\frac{\theta(\mathbf{x},t)}{2} \bm\nabla \theta(\mathbf{x},t)
\right]^2
\right\}
\nonumber \\
&=&
\frac{1}{4}
\left\{
\frac{1}{P(\mathbf{x},t)}\bigr[\bm\nabla P(\mathbf{x},t)\bigr]^2
+
P(\mathbf{x},t)\left[\bm\nabla \theta(\mathbf{x},t)\right]^2
\right\}
.
\label{PA10}
\end{eqnarray}
Similarly,
\begin{eqnarray}
\phi(\mathbf{x},t)\Phi^\dagger\Phi&=&\phi(\mathbf{x},t)P(\mathbf{x},t)
,
\\
\Phi^\dagger\sigma^x \Phi&=&P(\mathbf{x},t)\sin\theta(\mathbf{x},t)\cos\frac{S_2(\mathbf{x},t)-S_1(\mathbf{x},t)}{\hbar},
\\
\Phi^\dagger\sigma^y \Phi&=&P(\mathbf{x},t)\sin\theta(\mathbf{x},t)\sin\frac{S_2(\mathbf{x},t)-S_1(\mathbf{x},t)}{\hbar},
\\
\Phi^\dagger\sigma^z \Phi&=&P(\mathbf{x},t)\cos\theta(\mathbf{x},t)
.
\label{PA11}
\end{eqnarray}
Next, we introduce
\begin{eqnarray}
S_1(\mathbf{x},t)&=&S(\mathbf{x},t)-\frac{\hbar\varphi(\mathbf{x},t)}{2}
,
\nonumber \\
S_2(\mathbf{x},t)&=&S(\mathbf{x},t)+\frac{\hbar\varphi(\mathbf{x},t)}{2}
,
\label{PA12}
\end{eqnarray}
and obtain
\begin{eqnarray}
\frac{i\hbar}{2}
\left(
\frac{\partial \Phi^\dagger }{\partial t}\Phi
-
\Phi^\dagger \frac{\partial \Phi}{\partial t}
\right)
&=&
\frac{\partial S(\mathbf{x},t)}{\partial t}
\left[P_1(\mathbf{x},t)+P_2(\mathbf{x},t)\right]
-
\frac{\hbar}{2}
\frac{\partial \varphi(\mathbf{x},t)}{\partial t}
\left[P_1(\mathbf{x},t)-P_2(\mathbf{x},t)\right]
\nonumber \\
&=&
\left[
\frac{\partial S(\mathbf{x},t)}{\partial t}
-
\frac{\hbar}{2}
\cos\theta(\mathbf{x},t)\frac{\partial \varphi(\mathbf{x},t)}{\partial t}
\right]
P(\mathbf{x},t)
,
\\
\sum_{k=1}^2
\bigg[
\bm\nabla S_k(\mathbf{x},t)-{q}\mathbf{A}(\mathbf{x},t)
\bigg]^2
P_k(\mathbf{x},t)
&=&
P(\mathbf{x},t)
\sum_{k=1}^2
\frac{1-(-1)^k\cos\theta(\mathbf{x},t)}{2}
\bigg[
\bm\nabla S(\mathbf{x},t)
-{q}\mathbf{A}(\mathbf{x},t)
+\frac{(-\hbar)^k}{2}
\bm\nabla \varphi(\mathbf{x},t)
\bigg]^2
\nonumber \\
&=&
\left\{
\bigr[
\bm\nabla S(\mathbf{x},t)
-{q}\mathbf{A}(\mathbf{x},t)
\bigr]^2
+
\frac{\hbar^2}{4}
\bigr[
\bm\nabla \varphi(\mathbf{x},t)
\bigr]^2
\right\}
P(\mathbf{x},t)
\nonumber \\
&&
-\hbar\cos\theta(\mathbf{x},t)
\bm\nabla \varphi(\mathbf{x},t)
\bigr[
\bm\nabla S(\mathbf{x},t)
-{q}\mathbf{A}(\mathbf{x},t)
\bigr]
P(\mathbf{x},t)
\\
\Phi^\dagger\sigma^x \Phi&=&P(\mathbf{x},t)\sin\theta(\mathbf{x},t)\cos\varphi(\mathbf{x},t),
\\
\Phi^\dagger\sigma^y \Phi&=&P(\mathbf{x},t)\sin\theta(\mathbf{x},t)\sin\varphi(\mathbf{x},t),
\\
\Phi^\dagger\sigma^z \Phi&=&P(\mathbf{x},t)\cos\theta(\mathbf{x},t)
.
\label{PA13}
\end{eqnarray}

Collecting all terms we find
\begin{eqnarray}
Q= \int d\mathbf{x} \, dt
&\bigg(&
\frac{\hbar^2}{8m}
\left\{
\frac{1}{P(\mathbf{x},t)}\left[\bm\nabla P(\mathbf{x},t)\right]^2
+
\left[\bm\nabla \theta(\mathbf{x},t)\right]^2
P(\mathbf{x},t)
\right\}
+
\left[
\frac{\partial S(\mathbf{x},t)}{\partial t}
-
\frac{\hbar}{2}
\cos\theta(\mathbf{x},t)\frac{\partial \varphi(\mathbf{x},t)}{\partial t}
\right]
P(\mathbf{x},t)
\nonumber \\
&&+
\frac{1}{2m}
\left\{
\bigr[
\bm\nabla S(\mathbf{x},t)
-{q}\mathbf{A}(\mathbf{x},t)
\bigr]^2
+
\frac{\hbar^2}{4}
\left[
\bm\nabla \varphi(\mathbf{x},t)
\right]^2
-\hbar\cos\theta(\mathbf{x},t)
\bm\nabla \varphi(\mathbf{x},t)
\bigr[
\bm\nabla S(\mathbf{x},t)
-{q}\mathbf{A}(\mathbf{x},t)
\bigr]
\right\}
P(\mathbf{x},t)
\nonumber \\
&&+q\phi(\mathbf{x},t)P(\mathbf{x},t)
-\frac{q\hbar}{2m}
\bigr[
B_x\sin\theta(\mathbf{x},t)\cos\varphi(\mathbf{x},t)
+B_y\sin\theta(\mathbf{x},t)\sin\varphi(\mathbf{x},t)
+B_z\cos\theta(\mathbf{x},t)
\bigr]
P(\mathbf{x},t)
\bigg)
.
\nonumber \\
\label{PA14}
\end{eqnarray}

\end{widetext}
\section{Classical mechanics of a magnetic moment}\label{MM}
For completeness, we collect some well-known facts about the classical mechanical description
of the rotational motion of a magnetic moment $\mathbf{M}=\mathbf{M}(t)=(M_x,M_y,M_z)^T$ which does not move
and interacts with a magnetic field $\mathbf{B}=\mathbf{B}(t)=(B_x,B_y,B_z)^T$~\cite{SCHI62}.
The motion of the magnetic moment is completely determined by the torque equation
\begin{eqnarray}
\frac{d\mathbf{M}}{dt}&=&\gamma\mathbf{M}\times\mathbf{B}
,
\label{MM0a}
\end{eqnarray}
where $\gamma$ is the gyromagnetic ratio.
In terms of components we have
\begin{eqnarray}
\frac{d M_x}{dt}&=&\gamma(M_yB_z-M_zB_y)
\nonumber \\
\frac{d M_y}{dt}&=&\gamma(M_zB_x-M_xB_z)
\nonumber \\
\frac{d M_z}{dt}&=&\gamma(M_xB_y-M_yB_x)
.
\label{MM0}
\end{eqnarray}
Assuming that the total magnetic moment $M_0=(M_x^2+M_y^2+M_z^2)^{1/2}$ is constant, we can write
$\mathbf{M}=M_0 \mathbf{m}$ where
\begin{eqnarray}
\nonumber \\
\mathbf{m}=(\cos\varphi\sin\theta,\sin\varphi\sin\theta,\cos\theta)^T
,
\label{MM1}
\end{eqnarray}
is a unit vector and $(\theta,\varphi)$ are its spherical coordinates.
The equations of motion of these coordinates read
\begin{eqnarray}
\frac{d\varphi}{dt}&=&\frac{d \arctan(m_y/m_x)}{dt}
\nonumber \\
&=&\frac{\gamma}{m_x^2+m_y^2}\left(m_x\frac{d m_y}{dt}-m_y\frac{d m_x}{dt}\right)
\nonumber \\
&=&\gamma\left[-B_z + \frac{m_z}{m_x^2+m_y^2}(m_xB_x+m_yB_y)\right]
\nonumber \\
&=&\gamma\left[-B_z + \frac{z}{\sqrt{1-z^2}}(B_x\cos\varphi+B_y\sin\varphi)\right]
,
\label{MM2}
\end{eqnarray}
where $z=\cos\theta$ and
\begin{eqnarray}
\frac{d z}{dt}&=&\gamma(m_xB_y-m_yB_x)
\nonumber \\
&=&\gamma\left[\sqrt{1-z^2}(-B_x\sin\varphi+B_y\cos\varphi)\right]
.
\label{MM3}
\end{eqnarray}
If we define a Hamiltonian $H_M$ by
\begin{eqnarray}
H_M&=&-\gamma \left( B_x\sin\theta\cos\varphi+B_y\sin\theta\sin\varphi+B_z\cos\theta\right)
\nonumber \\
&=&-\gamma \left[ z B_z+ \sqrt{1-z^2}(B_x\cos\varphi+B_y\sin\varphi)\right]
,
\label{MM4}
\end{eqnarray}
it follows that
\begin{eqnarray}
\frac{d\varphi}{dt}&=&+\frac{\partial H_M}{\partial z}
\nonumber \\
\nonumber \\
\frac{d z}{dt}&=&-\frac{\partial H_M}{\partial \varphi}
.
\label{MM5}
\end{eqnarray}
From Eq.~(\ref{MM5}), it follows that the pair $(\varphi,z)$ are conjugate variables.

Finally, it is easy to check that the equations of motion Eq.~(\ref{MM5})
follow by searching for the extremum of the functional
\begin{eqnarray}
{\cal M}&=&\int dt\;
\left(
-z\frac{d\varphi}{dt}+H_M\right)
.
\label{MM6}
\end{eqnarray}

\bibliography{../../../all13}
\end{document}